\documentstyle[preprint,tighten,eqsecnum,aps]{revtex}

\begin{document}
\draft
\preprint{WU B 95-09}
\title{VIRTUAL COMPTON SCATTERING OFF PROTONS \\
AT MODERATELY LARGE MOMENTUM TRANSFER}
\author{P. Kroll\footnote
{kroll@wpts0.physik.uni-wuppertal.de}, M. Sch\"urmann}
\address{Fachbereich Physik, Universit\"at Wuppertal, \\ 
Gau\ss{}strasse 20, D-42097 Wuppertal, Germany}
\author{P.~A.~M.~Guichon\footnote{pampam@phnx7.saclay.cea.fr}}
\address{SPhN-DAPNIA, CE Saclay,\\
F-91191 Gif sur Yvette, France}
\date{\today}
\maketitle
\begin{abstract}
The amplitudes for virtual Compton scattering off protons are calculated 
within the framework of the diquark model in which protons are viewed as 
being built up by quarks and diquarks. The latter objects are treated as 
quasi-elementary constituents of the proton. Virtual Compton scattering, 
electroproduction of photons  and the Bethe-Heitler contamination are
discussed for various kinematical situations. We particularly emphasize
the r\^ole of the electron asymmetry for measuring the relative phases
between the virtual Compton and the Bethe-Heitler amplitudes. 
It is also shown that the model is able to describe very well the 
experimental data for real Compton scattering off protons. 
\end{abstract}
\narrowtext
\section{INTRODUCTION}
The understanding of the structure of the nucleon is a fundamental task 
of particle physics. Instead of inclusive reactions where the various 
nucleon constituents contribute incoherently, one may use exclusive 
reactions to study the structure of the nucleon. The simplest exclusive 
quantities are the electromagnetic form factors of the nucleon, which can 
be measured in elastic electron-proton and/or electron-deuteron scattering. 
The cross section for unpolarized elastic electron-proton scattering is well 
known up to a momentum transfer $Q^2$ of $33\,\mbox{GeV}^2$ \cite{Arn:88}. 
For large momentum transfer $(\geq 8.8\,\mbox{GeV}^2)$ only the magnetic 
form factor of the proton has been determined because the contribution of 
the electric one is suppressed in the cross section by $1/Q^2$. 
In a recent SLAC experiment \cite{Bos:92} the electric and magnetic form 
factors of the proton have been measured separately in the $Q^2$ range from 
$1.75\,\mbox{GeV}^2$ to $8.83\,\mbox{GeV}^2$. The data on the neutron form 
factors is very poor at large momentum transfer. 

The next simplest process to probe the structure of the nucleon 
is Compton scattering off protons at large momentum transfers. The cross 
section for real Compton scattering has been measured up to initial 
photon energies of $6\,\mbox{GeV}$ \cite{Shu:79}. The process 
$ep\rightarrow e p \gamma $ offers the possibility to study virtual Compton 
(VC) scattering and its interference with the Bethe-Heitler (BH) process. At 
present there is no experimental data available. In the near future, however, 
CEBAF will offer the possibility to measure the process 
$ep\rightarrow e p \gamma$ for incident electron energies smaller than 
$6\,\mbox{GeV}$ \cite{aud:93}. The investigation of the high energy region 
requires a new type of electron accelerator \cite{arv:93}. We note that the 
BH contribution to the procces $ep\rightarrow e p \gamma$ is well known 
\cite{Mo:69}.
 
The general framework for calculating large momentum transfer processes 
has been developed by Brodsky and Lepage \cite{Bro:80}. Using the QCD 
factorization theorem for exclusive reactions \cite{Bro:80,Mue:81}, one can 
write baryon form factors and Compton scattering off baryons as 
multidimensional integrals over a product of distribution amplitudes (DA) and 
elementary amplitudes which represent the scattering of constituents in 
collinear approximation. A DA is the valence Fock-state wave function 
integrated over transverse momenta ${\bf k}_\perp^{(i)}$ up to a scale 
of order $Q^2$. It specifies the distribution of the longitudinal momentum 
fractions the constituents carry inside their parent hadrons. 
Contributions from higher Fock states are suppressed by powers of  
$\alpha_s /Q^2$ in exclusive processes where $\alpha_s$ is the strong coupling
constant. 

Some information about DAs containing the non-perturbative 
physics, is obtained from QCD sum rules \cite{Che:84a,Che:84b,Kin:87} 
and from lattice gauge theory \cite{Kron:85,Ric:87}. The absolute normalization 
of the nucleon wave function has been estimated: QCD sum rules provide a 
value of $f_N=(5.0\pm 0.3)\cdot 10^{-3} \mbox{GeV}^2$ for the wave function at 
the origin of the configuration space whereas $f_N=(2.9\pm 0.6)\cdot 10^{-3} 
\mbox{GeV}^2$ has been obtained from lattice gauge theory. Also the moments of 
the nucleon's DA have been calculated up to order three. These moments are used
to constrain the first five coefficients in an expansion of the DA over the 
eigenfunctions of the evolution equation which are linear combinations of the 
Appell polynomials. It is obvious that the DAs do not unambiguously follow from
the few error-burdened moments provided by QCD sum rules or lattice gauge 
theories (a careful analysis of this problem has been performed by Bergmann and
Stefanis \cite{Ber:93}). Therefore, the DAs to be found in the literature, 
e.~g., \cite{Che:84b,Kin:87,Ber:93} are to be considered as models. Anyway,
the striking feature of these DAs is their strong asymmetry which 
gives preference to positive helicity $u$ quarks in positive helicity protons. 
The asymmetry of these DAs is the source of theoretical inconsistencies in 
applications of the Brodsky-Lepage model, because it strongly enhances the 
contributions from the end-point regions where $\alpha_s$ is large and, hence, 
the use of perturbation theory is unjustified. Recently it has been shown 
\cite{Li:92a,Li:92b,Bol:94} that perturbative calculations become theoretically
self-consistent for momentum transfers as low as a few $\mbox{GeV}^2$ when the 
transverse separations of the constituents as well as a Sudakov factor, 
comprising gluonic radiative corrections, are taken into account. If one also 
includes the intrinsic transverse momentum dependence of the hadronic wave 
function perturbative calculations may become self-consistent even for momentum 
transfers as low as $1$ to $2$ $\mbox{GeV}^2$ in the case of the pion 
\cite {Jak:93a} and at about $7$-$9\,{\rm GeV}^2$ for the nucleon 
\cite{Bol:94}. On the other hand, the inclusion of the transverse degrees of 
freedom and of the Sudakov factor leads to substantial suppressions of 
the perturbative contributions in the few GeV region which are particularly
strong for the asymmetric DAs.

To leading order perturbation theory many processes 
have been calculated within the standard approach of Brodsky and Lepage 
\cite{Bro:80}, among them is the pion's form factor, 
the magnetic form factor of the nucleon, two-photon annihilation into meson 
pairs and into proton-antiproton as well as  
Compton scattering off protons. To next to leading order in $\alpha_s$ only the
pion's form factor and two photon annihilation into meson pairs have been 
studied. It was shown \cite{Niz:87} that the corrections to the lowest order 
predictions for $\gamma\gamma\rightarrow M\bar{M}$ become sufficiently 
small $(\le 25\%)$ only for very large center of mass energies 
$\sqrt{s} \le 10\,\mbox{GeV}$. Other important reactions like elastic 
proton-proton scattering seem to be beyond feasibility with present-day 
techniques of computing very large numbers of Feynman diagrams. There are also 
unsolved theoretical complications with pinch singularities \cite{Mue:81}.

There are two characteristic features of the Brodsky-Lepage model, 
commonly termed the hard scattering approach (HSA): the dimensional 
counting rules and the conservation of hadronic helicities. 
The latter feature implies that any helicity flip amplitude is zero and, hence,
any single spin asymmetry too. The helicity sum rule is a consequence of 
utilizing the collinear approximation and of dealing with (almost) massless 
quarks which conserve their helicities when interacting with gluons. Whereas 
the dimensional counting rules are in reasonable agreement with experiment, 
seems the helicity sum rule to be violated by $20-25\%$. For instance, the 
polarization parameter $P$ in elastic proton-proton scattering is by no means 
zero. On the contrary, it is rather large $(20-30\%)$ at 
moderately large values of momentum transfer $(p_{T}>2GeV/c)$ \cite{Cra:90}. 
Even worse, the data show the tendency of further rise of the polarization with
increasing $p_{T}$. Such peculiar polarization phenomena have also been 
observed in many (moderately) large $p_{T}$ inclusive reactions like hyperon 
production in nucleon-nucleon collisions or $pp\rightarrow\pi^{\pm}X$ 
\cite{Sar:90}. The prevailing opinion is that also these phenomena cannot 
be explained in terms of perturbative QCD (see, for example, 
Ref.~\cite{Kro:88,Siv:89}), rather they are generated by an interplay of 
perturbative and non-perturbative physics.

A model which takes into considerations non-perturbative effects, has been 
proposed by us in a series of papers \cite{Ans:87,Kro:87,Kro:93,Jak:93b}. 
In that model baryons are viewed as made of quarks and diquarks, the latter 
being treated as quasi-elementary constituents which partly survive medium 
hard collisions. The composite nature of the diquarks is taken into account by 
diquark form factors. Diquarks are an effective description of correlations in 
the wave functions and constitute a particular model for non-perturbative 
effects. The diquark model has been applied to a variety of processes and 
successfully confronted to data. Among these applications is a recent study of 
the nucleon's electromagnetic form factors \cite{Jak:93b}. In fact, predictions
are achieved for both the magnetic and the electric form factors. For the 
latter quantity no result is obtained in the pure quark hard scattering model 
because it requires helicity flips of the nucleon and, in so far, the electric 
form factor or better the Pauli form factor, also represents a polarization 
effect. In the diquark model helicity flips are generated through spin 1 (V) 
diquarks. The diquark model is designed such that it turns into the 
theoretically well established pure quark picture asymptotically. In so far
the pure quark picture of Brodsky-Lepage and the diquark model do not 
oppose each other, they are not alternatives but rather complements. 

The purpose of this paper is to discuss virtual Compton scattering 
off protons and its interference with the Bethe-Heitler process where the 
photon is emitted by the electrons (see Fig.~1). The Compton amplitudes are 
calculated within the framework of the diquark model with the parameters taken 
from previous studies \cite{Kro:93,Jak:93b}. Real and virtual Compton 
scattering are the two simplest experimentally accessible processes in which 
the integrals over the longitudinally momentum fractions yield imaginary parts.
The reason is that there are kinematical regions where internal quarks and 
gluons can go on mass shell. It was shown in \cite{Far:89a} 
that the lowest order predictions of perturbative QCD for photon induced 
processes need no resummation of Sudakov corrections. The essential point is 
that the singularities do not pinch as is the case in baryon-baryon scattering.
The appearance of imaginary parts to leading order in $\alpha_s$ is a 
non-trivial prediction of perturbative QCD. The process $ep\rightarrow e p 
\gamma $ with polarized incoming electrons offers the possibility to determine 
the phases of the Compton amplitudes as a function of the center of mass 
scattering angle, by investigating the interference of the virtual Compton 
with the Bethe-Heitler process. In principle, the measurements of other spin 
observables through the use of polarized targets and polarized beams would 
offer further possibilities of crucially testing the applicability of the HSA 
in the few GeV region. The process $ep\rightarrow e p \gamma$ has been studied
in the framework of the pure quark HSA by Farrar and Zhang \cite{Far:89}. 
Real Compton scattering has, on the other side been investigated by two groups 
\cite{Far:89,Kro:91}. Unfortunately, the results obtained by the 
two groups deviate from each other substantially. It is suspected that the 
discrepancies are caused by an improper treatment of propagator singularities 
in \cite{Far:89}. The kinematical flexibility of $ep\rightarrow e p \gamma $ 
allows to test the predictions with respect to the virtual photon mass, to the 
center of mass energy and to the momentum transfer. 

In Sect.~\ref{skin} a detailed discussion of the kinematics and 
the various observables of the reaction $ep\rightarrow e p \gamma$ is given. 
A short description of the diquark model for exclusive reactions is 
presented in Sect.~\ref{sdiq} followed by a brief discussion of the 
results for real Compton scattering (Sect.~\ref{srcs}). The predictions for 
the cross section of VC scattering and for the unpolarized
cross section of the photon electroproduction $ep\rightarrow e p \gamma$ are 
discussed in Sect.~\ref{sr}. This section also includes a discussion of the 
interference between the VC amplitudes and the BH ones. The case of polarized 
electron beams and its physics implications is investigated in Sect.~6. The 
paper terminates with a few concluding remarks (Sect.~7).

\section{Kinematics of electroproduction of photons}
\label{skin}
The process $e\,p\rightarrow e\,p\,\gamma$ receives contributions 
from the VC and from the BH processes. The kinematics and the helicities are 
specified in Fig.~\ref{fkin}. We work in the center of mass (CM) frame of the 
final state photon and proton. Neglecting the electron mass the 
different particle momenta can be chosen to be
\begin{eqnarray}
&&k^\mu=\left(k_0,\,k_0\sin\alpha\cos\phi,\,k_0\sin\alpha\sin\phi,\,
                    k_0\cos\alpha\right) \nonumber\\
&&k'^\mu=\left(k_0-q_0,k_0\sin\alpha\cos\phi,\,k_0\sin\alpha\sin\phi,\,
                    k_0\cos\alpha-|{\bf p}|\right) \nonumber\\
&&q^\mu=\left(q_0,\,0,\,0,\,|{\bf p}|\right) \nonumber\\
&&q'^\mu=q_0'\left(1,\,\sin\theta,\,0,\,\cos\theta\right) \nonumber\\
&&p^\mu=\left(p_0,\,0,\,0,\,-|{\bf p}|\right) \nonumber\\
&&p'^\mu=\left(p_0',\,-q_0'\sin\theta,\,0,\,-q_0'\cos\theta\right) 
\end{eqnarray}
in that frame. $\phi$ is the angle between the hadronic and leptonic scattering
plane, and $\theta$ is the scattering angle of the outgoing real photon. The 
polarization vector $\epsilon_f$ of the outgoing photon is given by
\begin{equation} 
\epsilon_f(\mu')=\frac{1}{\sqrt{2}}\left(0,\,-\mu'\cos\theta,\,-i,\,
                                        \mu'\sin\theta\right).
\end{equation}
$q_0,\,p_0,\,q_0',\,p_0'$ and $|{\bf p}|$ are related to the Mandelstam variable $s=(p+q)^2$ and to the virtuality ($q^2=-Q^2$) of the exchanged photon by 
\begin{eqnarray}
 q_0=\frac{s-Q^2-m^2}{2\sqrt{s}}\qquad\qquad q_0'=\frac{s-m^2}{2\sqrt{s}}
                                     \nonumber \\
 p_0=\frac{s+Q^2+m^2}{2\sqrt{s}} \qquad\qquad p_0'=\frac{s+m^2}{2\sqrt{s}}  
                                     \nonumber \\ 
|{\bf p}|=\frac{1}{2\sqrt{s}}\,\Lambda(s,\,-Q^2,\,m^2)
\end{eqnarray}
where $m$ is the nucleon mass. The Mandelstam function $\Lambda$ is defined by
\begin{equation}
\Lambda(x,y,z)=\sqrt{x^2+y^2+z^2-2\,x\,y-2\,x\,z-2\,y\,z}.
\end{equation}
Momentum conservation leads to the following relations
\begin{equation}
k_0\sin\alpha=\frac{Q}{2|{\bf p}|}\sqrt{(2\,k_0-q_0)^2-|{\bf p}|^2}, 
\hspace{1.5cm}
k_0\cos\alpha=\frac{1}{2|{\bf p}|}\left(q_0\,(2\,k_0-q_0)+|{\bf p}|^2\right).\\
\end{equation}
The other frequently used reference frame is the laboratory frame 
defined by ${\bf p}=0$. In that frame one has the following 
useful relations:
\begin{eqnarray}
&&s=-Q^2+m^2+2\,m\,k_{0L} \nonumber\\
&&Q^2=4\,k_{0L}k_{0L}'\sin^2\frac{\theta_e^L}{2}
\end{eqnarray}
where $k_{0L}$ and $k_{0L}'$ are the energies of the inital and final state 
electrons in the LAB and $\theta_e^L$ is the electron scattering angle. 
>From the theoretical point of view the most significant variables 
are the invariants $s$, $t$, $Q^2$ and the polarization of the virtual photon 
in the VC process   
\begin{equation}
\varepsilon:=\frac{(q_0-2\,k_0)^2-|{\bf p}|^2}{(q_0-2\,k_0)^2+|{\bf p}|^2} 
=\left[1+2\,\frac{(k_{0L}-k_{0L}')^2+Q^2}{Q^2}
\tan^2{\frac{\theta_e^L}{2}}\right]^{-1}.
\end{equation}
The unpolarized differential cross section for the reaction 
$e p\rightarrow e p\gamma$ can be written as
\begin{equation}
\label{dcsep}
\frac{d^4\sigma}{ds\, dQ^2\, d\phi\, dt}(e p\rightarrow e p\gamma)=
\frac{1}{128\,(2\pi)^4\,k_{0L}^2\,m^2\,\Lambda(s,-Q^2,m^2)}\frac{1}{4}
\sum_{\nu\,\nu'\atop\mu'\lambda'\lambda}
\left|T^{BH}_{\nu'\mu'\lambda',\nu\lambda}
     +T^{VC}_{\nu'\mu'\lambda',\nu\lambda}
\right|^2
\end{equation}
where $T^{BH}$ and $T^{VC}$ denote the helicity amplitudes for the BH and for 
the VC contributions to the electroproduction of photons, respectively. In the 
one-photon approximation the latter read (see Fig.~\ref{fkin})
\begin{equation}
\label{vca1}
T^{VC}_{\nu'\mu'\lambda',\nu\lambda}=\frac{\sqrt{4\pi\alpha}}{Q^2}
\delta_{\nu'\nu}\, \bar{u}(k',\nu')\gamma^\alpha u(k,\nu)\,
\langle q'\mu',p'\lambda'|j_\alpha^{VC}|p\lambda\rangle .
\end{equation}
In the CM frame the two-body $s$-channel helicity amplitudes 
$\Phi_i:=M_{\mu'\lambda',\mu\lambda}(s,t,Q^2)$ for VC scattering are defined by
\begin{equation}
M_{\mu'\lambda',\mu\lambda}=\epsilon^\alpha(\mu)\,
\langle q'\mu',p'\lambda'|j_\alpha^{VC}|p\lambda\rangle
\end{equation}
where $\epsilon_i^\alpha$ is the polarization vector of the virtual photon 
\begin{equation}
\epsilon_i^\alpha(0)=\frac{1}{Q}\left(|{\bf p}|,0,0,q_0\right)
\hspace{1cm}
\epsilon_i^\alpha(\pm 1)=\mp\frac{1}{\sqrt{2}}\left(0,1,i,0\right) .
\end{equation}
Current conservation leads to a relation between the 0-and 3-components 
of the hadronic current matrix element. The projection of the electron 
current on the polarization vectors is straightforward and leads to the 
following relation between the VC contribution to the photon electroproduction 
amplitudes and the VC helicity amplitudes 
\begin{eqnarray}
\label{tvc}
T^{VC}_{\nu'\mu'\lambda',\nu\lambda}
=\frac{1}{Q}\sqrt{\frac{4\pi\alpha}{1-\varepsilon}}
\left\{
\frac{1}{\sqrt{2}}\left[\sqrt{1+\varepsilon}+2\nu\,\sqrt{1-\varepsilon}\right]
e^{-i\phi}M_{\mu'\lambda',+1\lambda} 
-\sqrt{2\,\varepsilon}\,M_{\mu'\lambda',0\lambda}
\right.  \nonumber\\
\hspace{3.0cm}
-\left.\frac{1}{\sqrt{2}}\left[\sqrt{1+\varepsilon}-2\nu\,\sqrt{1-\varepsilon}
\right]
e^{i\phi}M_{\mu'\lambda',-1\lambda})\right\}\delta_{\nu'\nu} .
\end{eqnarray}
Due to parity invariance there are only 12 independent $s$-channel helicity amplitudes contributing to the process 
$\gamma^\ast\,p\rightarrow\gamma\,p$. These are conveniently denoted by 
$\Phi_i$, $i=1,\,12$:  
\begin{equation}
\label{helcs}
\begin{array}{lll}
  \Phi_{1}=M_{+1+\frac{1}{2},1+\frac{1}{2}} \qquad\qquad
& \Phi_{5}=M_{+1-\frac{1}{2},1-\frac{1}{2}} \qquad\qquad
& \Phi_{9}=M_{+1+\frac{1}{2},0+\frac{1}{2}} \\
  \Phi_{2}=M_{-1-\frac{1}{2},1+\frac{1}{2}} \qquad\qquad
& \Phi_{6}=M_{-1+\frac{1}{2},1-\frac{1}{2}} \qquad\qquad
& \Phi_{10}=M_{-1-\frac{1}{2},0+\frac{1}{2}} \\
  \Phi_{3}=M_{-1+\frac{1}{2},1+\frac{1}{2}}  \qquad\qquad
& \Phi_{7}=M_{-1-\frac{1}{2},1-\frac{1}{2}}  \qquad\qquad
& \Phi_{11}=M_{-1+\frac{1}{2},0+\frac{1}{2}} \\
  \Phi_{4}=M_{+1-\frac{1}{2},1+\frac{1}{2}} \qquad\qquad
& \Phi_{8}=M_{+1+\frac{1}{2},1-\frac{1}{2}} \qquad\qquad
& \Phi_{12}=M_{+1-\frac{1}{2},0+\frac{1}{2}} 
\end{array}
\end{equation}
$\Phi_1-\Phi_8$ correspond to transverse polarization of the virtual photon 
and $\Phi_9-\Phi_{12}$ to longitudinally polarized photons. In the 
real photon limit the latter four amplitudes vanish. In that limit time 
reversal invariance reduces the number of independent amplitudes even further: 
$(\Phi_7=\Phi_3,\,\Phi_8=-\Phi_4)$. The VC contribution to the 
$ep\to ep\gamma$ cross-section can be decomposed as follows 
\begin{eqnarray}
\label{d4svc}
\left. \frac{d^4\sigma}{ds\, dQ^2\, d\phi\, dt}\right|_{VC}=
\frac{\alpha (s-m^2)}{4\,(2\pi)^2\,k_{0L}^2\,m^2\,Q^2\,(1-\varepsilon)}
\hspace{3.0cm}\nonumber \\
\times\left(\frac{d\sigma_T}{dt}+\varepsilon\,\frac{d\sigma_L}{dt} 
+\varepsilon\cos{2\phi}\,\frac{d\sigma_{TT}}{dt}
+\sqrt{2\varepsilon(1+\varepsilon)}\cos{\phi}\,\frac{d\sigma_{LT}}{dt}\right)
\end{eqnarray}
where the partial cross-sections read:\\
i) The cross-section for transverse photons (which, at $Q^2=0$, reduces to the 
unpolarized cross section $d\sigma/dt$ for real Compton scattering)
\begin{equation}
\label{sigt}
\frac{d\sigma_T}{dt}=\frac{c}{2}\;\sum_{i=1}^{8}\; |\Phi_i|^2.
\end{equation}
ii) The cross-section for longitudinal photons 
\begin{equation}
\label{sigl}
\frac{d\sigma_L}{dt}=c\;\sum_{i=9}^{12}\; |\Phi_i|^2.
\end{equation}
iii) The transverse-transverse interference term
\begin{equation}
\label{sigtt}
\frac{d\sigma_{TT}}{dt}=
-\frac{c}{2}\;\Re e\left[\Phi_1^\ast\,\Phi_7-\Phi_2^\ast\,\Phi_8
                     +\Phi_3^\ast\,\Phi_5-\Phi_4^\ast\,\Phi_6\right].
\end{equation}
iv) The longitudinal-transverse interference term
\begin{equation}
\label{siglt}
\frac{d\sigma_{LT}}{dt}=-\frac{c}{\sqrt{2}}
\;\Re e\left[\Phi_9^\ast\,(\Phi_1-\Phi_7)
      +\Phi_{10}^\ast\,(\Phi_2+\Phi_8)
      +\Phi_{11}^\ast\,(\Phi_3-\Phi_5)
      +\Phi_{12}^\ast\,(\Phi_4+\Phi_6)\right].
\end{equation}
The phase space factor $c$ is given by
\begin{equation}
c=\frac{1}{16\pi (s-m^2) \Lambda(s,-Q^2,m^2)}
\end{equation}
The Bethe-Heitler amplitudes read (see Fig. \ref{fkin})
\begin{equation}
\label{bh1}
T^{BH}_{\nu'\mu'\lambda',\nu\lambda}=\frac{4\pi\alpha}{t}\,L_\alpha^{BH}
\langle p'\lambda'|j^\alpha|p\lambda\rangle
\end{equation}
where the $\gamma^\ast\,p\rightarrow p$ current matrix element is expressed in 
terms of the magnetic form factor of the proton $G_M$ and the Pauli form 
factor $F_2$ by:
\begin{equation}
\langle p'\lambda'|j^\alpha|p\lambda\rangle
=\sqrt{4\pi\alpha}\,\bar{u}(p',\lambda')\left(\gamma^\alpha G_M(t)
-\frac{\kappa_p}{2\,m}(p'+p)^\alpha F_2(t)\right) u(p,\lambda)\,.
\end{equation}
The electric form factor $G_E$ is related to $F_2$ by 
$G_E=G_M-\kappa_p(1+\tau)F_2$ where $\tau=-t/4m^2$.
The leptonic current $L_\alpha^{BH}$ is given by
\begin{eqnarray}
\label{lcbh}
L_\alpha^{BH}&=\left\{2\frac{k'\cdot \epsilon_f^\ast}{s_{k'q'}}
+2\frac{k \cdot \epsilon_f^\ast}{t_{kq'}}\right\}
\bar{u}(k',\nu')\gamma^\alpha u(k,\nu) \nonumber\\
&&+\bar{u}(k',\nu')\left\{   \frac{\epsilon_f^\ast\hspace{-0.35cm}/\;q'
\hspace{-0.3cm}/\;\gamma_\alpha}{s_{k'q'}}  -\frac{\gamma_\alpha\;q'
\hspace{-0.3cm}/\;\epsilon_f^\ast\hspace{-0.35cm}/}{t_{kq'}}
\right\} u(k,\nu) 
\end{eqnarray}
where $s_{ab}=(a+b)^2$ and $t_{ab}=(a-b)^2$. In the soft photon approximation 
$q'\to 0$, the first term in (\ref{lcbh}) is the usual Bremsstrahlung 
contribution. Putting all together the BH contribution to the process 
$e p\to e p\gamma$ may be written in a form similar to that of the Rosenbluth 
cross section for elastic $ep$ scattering:
\begin{eqnarray}
\label{d4sbh}
\lefteqn{
\left. \frac{d^4\sigma}{ds\, dQ^2\, d\phi\, dt}\right|_{BH}=
\frac{\alpha^3}{4\,\pi\,k_{0L}^2\,m^2\,\Lambda(s,-Q^2,m^2)\,t^2}}\\
&&\times\left\{A(s_{kp},s,Q^2,\phi,t)\frac{G_E^2(t)+\tau G_M^2(t)}{1+\tau}
+B(s_{kp},s,Q^2,\phi,t)\,G_M^2(t)\right\}.
\end{eqnarray}
The functions $A$ and $B$ are given by
\begin{eqnarray}
&&A(s_{kp},s,Q^2,\phi,t)=m^2\left(\frac{s_{k'q'}}{t_{kq'}}
+\frac{t_{kq'}}{s_{k'q'}}\right)
+\frac{t}{s_{k'q'}}\left(s_{kp}-s-Q^2-2\,m^2\right)\\
&&-\frac{t}{t_{kq'}}\left(s_{kp}+m^2\right)\nonumber
+\frac{t}{s_{k'q'}\,t_{kq'}}\left(2\,(s_{kp}-m^2)(s_{kp}-s-Q^2)
-u(s+Q^2)+m^2(t+m^2)\right)\,,\\
&&B(s_{kp},s,Q^2,\phi,t)=\frac{t}{2}\left(\frac{s_{k'q'}}{t_{kq'}}
+\frac{t_{kq'}}{s_{k'q'}}\right)-\frac{t^2\,Q^2}{s_{k'q'}\,t_{kq'}}\,.
\end{eqnarray}
For large t the contribution from the electric form factor is suppressed by 
the factor $\tau$ as compared to that from the magnetic one. The interference 
term between the BH and the VC contribution is a very lengthy expression and 
we refrain from showing it.\\
Similar expressions as (\ref{d4svc}) and (\ref{d4sbh}) hold also for polarized 
cross sections. However, most likely only the polarization of the electron 
beams can be used at high energies. Therefore, we refain from discussing 
polarized cross sections in this article with the only exceptions of the 
electron asymmetry $A_L$ obtained by reversing the helicity of the beam 
electrons, and the corresponding proton asymmetry. The discussion of these 
asymmetries is postponed to Sect.~6.

\section{The diquark model}
\label{sdiq}
In the hard-scattering model of Brodsky-Lepage \cite{Bro:80} 
the process $\gamma^\ast\,p\rightarrow\gamma\,p$
is expressed by a convolution of DAs with hard-scattering amplitudes 
calculated in collinear approximation within perturbative QCD.
In a collinear situation in which intrinsic transverse momenta
are neglected and all constituents of a hadron have momenta
parallel to each other and parallel to the momentum of the parent
hadron, one may write the valence Fock state of the proton in a covariant 
fashion (omitting colour indices for convenience)
\begin{equation}
\label{pwf}
|p,\lambda\rangle  = f_S\,\varphi_S(x_1)\,B_S\, u(p,\lambda) 
             + f_V\, \varphi_V(x_1)\, B_V
              (\gamma^{\alpha}+p^{\alpha}/m)\gamma_5 \,u(p,\lambda)/\sqrt{3}
\end{equation}
$u$ is the spinor of a proton with momentum $p$ and helicity $\lambda$.
The two terms in (\ref{pwf}) represent configurations 
consisting of a quark and either a spin-isospin zero $(S)$ or a 
spin-isospin one $(V)$ diquark, respectively. The couplings of the diquarks 
with the quarks in a proton lead to the flavour functions
\begin{equation}
\label{fwf}
B_S=u\, S_{[u,d]}\hspace{2cm} 
B_V= [ u V_{\{u,d\}} -\sqrt{2} d\, V_{\{u,u\}}]/\sqrt{3}\, .
\end{equation}
The use of covariant spin wave functions has many technical advantages.
For instance the calculation of a large set of so-called
elementary amplitudes is avoided, one immediately projects onto
hadronic states. Another advantage is that only hadronic quantities
(spinors, polarization vectors, and so on) appear. 
The DA $\varphi_{S(V)}(x_1)$, where $x_1$ is the momentum fraction carried 
by the quark, represents the light-cone wave function integrated over 
transverse momentum and is defined in such way that
\begin{equation}
\label{DAn}
\int_0^1 dx_1\,\varphi_{S,(V)}(x_1)=1\;.
\end{equation}
The constant $f_{S(V)}$ acts as the value of the configuration space 
wave function at the origin. \\
Representative Feynman graphs contributing to the hard-scattering 
amplitudes for the process of interest in this article, are displayed in 
Fig.~\ref{frfd}. The blobs appearing at the $gD$, $\gamma gD$ and
$\gamma\gamma D$ vertices symbolize three-, four- and five-point 
functions. These n-point functions are evaluated for point-like diquarks 
and the results are multiplied with phenomenological vertex 
functions (diquark form factors) which take into account the 
composite nature of the diquarks. Admittedly, that recipe is a rather crude 
approximation for $n\ge4$. Since the contributions from the n-point functions 
for $n\ge 4$ only represent small corrections to the final results that recipe 
is perhaps sufficiently accurate. The perturbative part of the model, i.e.~the 
coupling of gluons (and photons) to diquarks follows standard prescriptions, 
see e.~g., \cite{Lee:62}. The diquark-gluon vertices read (refer to 
\cite{Kro:87,Kro:93} for notations)
\begin{eqnarray}
&& \mbox{SgS}: i\,g_s t_{ij}^{a}\,(p_1+p_2)_{\mu} \nonumber\\
&& \mbox{VgV}: -i\,g_{s}t_{ij}^{a}\, 
\left\{
 g_{\alpha\beta}(p_1+p_2)_{\mu}
- g_{\mu\alpha}\left[(1+\kappa)\,p_1-\kappa\, p_2\right]_{\beta}
- g_{\mu\beta} \left[(1+\kappa)\,p_2-\kappa\, p_1\right]_{\alpha}
\right\}
\end{eqnarray}
where $g_s=\sqrt{4\pi\alpha_s}$ is the QCD coupling constant.
$\kappa$ is the anomalous magnetic moment of the vector diquark and 
$t^a=\lambda^a/2$ the Gell-Mann colour matrix. For the coupling of 
photons to diquarks one has to replace $g_s t^a$ by $-\sqrt{4\pi\alpha} e_D$ 
where $\alpha$ is the fine structure constant and $e_D$ is the electrical 
charge of the diquark in units of the elementary charge. The couplings 
$DgD$ are supplemented by appropriate contact terms required by 
gauge invariance
\begin{eqnarray}
&& \mbox{$\gamma$SgS}:
-2 \,i\,e_0 e_S g_s t^a_{ij}\, g_{\mu\nu} \nonumber\\
&& \mbox{$\gamma$VgV}:
i \,e_0 e_V g_s t^a_{ij} 
\left(2\,g_{\mu\nu}g_{\alpha\beta}
-g_{\mu\beta}g_{\alpha\nu}-g_{\mu\alpha}g_{\beta\nu}\right)
\end{eqnarray}
As we already mentioned the composite nature of the diquarks is taken into 
account by phenomenological vertex functions. Advice for the parameterization 
of the 3-point functions, ordinary  diquark form factors, is 
obtained from the requirement that asymptotically the diquark 
model evolves into the hard scattering model of Brodsky-Lepage. 
This requirement fixes the asymptotic 
behaviour of the form factors. Interpolating smoothly  between 
that behaviour and the conventional value of 1 at $Q^{2}=0$, 
the form factors are actually parametrized as
\begin{eqnarray}
\label{fs3}
F_{S}^{(3)}(Q^{2})=\frac{Q_{S}^{2}}{Q_{S}^{2}+Q^{2}}\,,\qquad
F_{V}^{(3)}(Q^{2})=\left(\frac{Q_{V}^{2}}{Q_{V}^{2}+Q^{2}}\right)^{2}\,.
\end{eqnarray}
The asymptotic behaviour of the diquark form factors and the connection to 
the hard scattering model is discussed in more detail in Ref.~\cite{Kro:87}.\\  
In accordance with the required asymptotic behaviour the $n$-point
functions for $n\geq 4$ are parametrized as
\begin{eqnarray}
\label{fsn}
F_{S}^{(n)}(Q^{2})=a_{S}F_{S}^{(3)}(Q^{2})\,,\qquad
F_{V}^{(n)}(Q^{2})=
\left(a_{V}\frac{Q_{V}^{2}}{Q_{V}^{2}+Q^{2}}\right)^{n-3}F_{V}^{(3)}(Q^{2}).
\end{eqnarray}
The constants $a_{S,V}$ are strength parameters. Indeed, since the diquarks in 
intermediate states are rather far off-shell one has to consider 
the possibility of diquark excitation and break-up. Both these possibilities 
would likely lead to inelastic reactions. Therefore, we have not to consider 
these possibilities explicitly in our approach but excitation and break-up 
lead to a certain amount of absorption which is taken into account by the 
strength parameters.\\ 
Actually, for our numerical studies we use \cite{Kro:93,Jak:93b}
\begin{equation}
\label{a10}
\begin{array}{l}
\varphi_S(x_1)=N_S\, x_1 x_2^3\exp{\left[-b^2 (m^2_q/x_1+m^2_S/x_2)\right]}\\
\varphi_V(x_1)=N_V\, x_1 x_2^3(1+5.8\,x_1-12.5\,x_1^2)
\exp{\left[-b^2 (m^2_q/x_1+m^2_V/x_2)\right]} 
\end{array}
\end{equation}
and the set of parameters
\begin{equation}
\label{c1}
\begin{array}{cccc}
 f_S= 73.85\,\mbox{MeV},& Q_S^2=3.22 \,\mbox{GeV}^2, & a_S=0.15, &  \\
 f_V=127.7\,\mbox{MeV},& Q^2_V=1.50\,\mbox{GeV}^2, & a_V=0.05,&\kappa=1.39\,;
\end{array}
\end{equation}
$\alpha_s=12\pi/25\,\log(Q^2/\Lambda_{QCD})$ is evaluated with 
$\Lambda_{QCD}=200\,\mbox{MeV}$ and restricted to be smaller than $0.5$. The 
parameters $Q_S$ and $Q_V$, controlling the size of the diquarks,
are in agreement with the higher-twist effects observed in the structure 
functions of deep inelastic lepton-hadron scattering \cite{Vir:91} if these 
effects are modeled as lepton-diquark elastic scattering. The DAs are
a kind of harmonic oscillator wave function transformed to the light cone.
The masses in the exponentials are constituent masses since they enter 
through a rest frame wave function. For the quarks we take $330 \,\mbox{MeV}$
whereas for the diquarks a value of $580 \,\mbox{MeV}$ is used. The oscillator 
parameter $b$ is taken to be $0.498\,\mbox{GeV}^{-1}$. The constant 
$(N_S=25.97;\,N_V=22.29)$ are fixed by the requirement (\ref{DAn}). The more 
complicated form of the DA $\varphi_V$ causes a smaller mean value of $x_1$ 
than obtained for the DA $\varphi_S$. The exponentials in the DAs provide 
strong suppressions in the end-point regions. 

According to the above rules the hard scattering amplitudes for the process 
$\gamma^\ast\,p\rightarrow\gamma\,p$ are calculated from the set of lowest 
order graphs of which a few representatives are shown in Fig.~\ref{frfd}. In 
fact, one has to evaluate 32 graphs for each of the diquarks, S and V. For 
comparison, in the pure quark model 366 graphs contribute. The calculation of 
the 64 graphs has been carried out using the SUN version of FORM \cite{Ver:90}.
We have checked that our amplitudes are gauge invariant with respect to the 
photons and the gluon. The $\gamma^\ast\,p\rightarrow\gamma\,p$ helicity 
amplitudes, decomposed in terms of the various n-point contributions, read  
\begin{eqnarray}
&&\Phi_i=\frac{(4\pi)^2\alpha}{9}C_F\int_0^1\,dy_1\int_0^1\,dx_1
\left\{f_S^2\,\varphi_S(y_1)
\left[4\,\Phi_i^{(S,3)}+2\,\Phi_i^{(S,4)}+\Phi_i^{(S,5)}\right]
\varphi_S(x_1)\right. \nonumber\\
&&\hspace{2cm}\left.
+f_V^2\,\varphi_V(y_1)
\left[2\,\Phi_i^{(V,3)}-2\,\Phi_i^{(V,4)}+11\,\Phi_i^{(V,5)}\right]
\varphi_V(x_1)\right\}
\end{eqnarray}
The factor $C_F$ ($=4/3$) is the Casimir operator of the fundamental 
representation of ${\rm SU}(3)_c$. The n-point contributions generated by the 
scalar diquarks are explicitely given in the Appendix \ref{aha}. The vector 
diquark contributions form very lengthy expressions and we refrain form quoting 
them here; they can be obtained from the authors on request.

\section{Results for real Compton scattering}
\label{srcs}
The results of the diquark model for real Compton scattering obtained 
from the DAs and the parameters in (\ref{a10}) and (\ref{c1}) are shown in 
Fig.~\ref{frcc} for three different initial photon energies. Note that
in the very forward and in the very backward regions the transverse momentum 
of the outgoing photon 
is small and, hence, our model which is based on perturbation theory, is not
applicable. As compared to the results presented by Kroll, Sch\"urmann and 
Schweiger \cite{Kro:87} there are small modifications of minor importance due 
to the use of covariant spin wave functions (\ref{pwf}) and to small changes 
of the parameters. For the vector diquark contributions the covariant spin wave
functions leads to correction terms related to helicity flips of the quarks.
Such terms have been neglected in previous work \cite{Kro:87}. 

Although experimental data are available only at energies which are at the 
limits of applicability of a model based on perturbative QCD, the diquark 
model is seen to work surprisingly well. The results obtained 
within the pure quark HSA by Kronfeld and ${\rm Ni\check{z}i\acute{c}}$ 
\cite{Kro:91} are of similar quality. It is interesting to see
that our predictions for the Compton cross section do not behave as 
$\sim s^{-6}$ at fixed angles and finite energies as the pure quark HSA
predicts. The reason for the deviations from the scaling law is obvious:
the various contributions to the cross section exhibit different energy
dependences due to the diquark form factors. The diquark model also 
predicts interesting photon asymmetries and spin correlation parameters
(see the discussion in \cite{Kro:87}). Even a transverse polarization
of the proton, of the order of $10\%$, is predicted \cite{Kro:87}. This 
comes about as a consequence of the perturbative phases of the amplitudes 
produced by the propagator poles and of non-zero helicity flip amplitudes
generated by the vector diquarks. In the pure quark perturbative approach
transverse asymmetries are zero (for massless quarks).    
\section{Cross sections for VC scattering and for electroproduction
of photons}
\label{sr}
In this section we present the predictions from the diquark model for
VC scattering. We select two CM energies, namely $s=5\;{\rm GeV}^2$ which
is the high energy end of the CEBAF accelerator (and is at the same time 
about the lower limit of the applicability of the diquark model) and 
a rather large energy of $s=10\;{\rm GeV}^2$ which may become accessible 
with future accelerators like, for instance, ELFE. 

In Fig.~\ref{fVC5} we show the $\gamma^\ast p \to \gamma p$ cross sections
for several values of $Q^2/s$. The transverse cross section (\ref{sigt})
for VC scattering is scaled by $s^6$ according to dimensional counting.
The most noticeable fact seen in the transverse cross section is the strong
decrease with $Q^2$, starting from $Q^2=0$, the real Compton scattering case.
The effect is particular dramatic in the backward direction. It can also be
observed from  Fig.~\ref{fVC5} that, as for real Compton scattering, the 
dimensional counting rule only holds approximately within a factor of 2
between 5 and $10\;{\rm GeV}^2$. This is a characteristic feature of the 
diquark model above-mentioned: the diquark form factors lead to a transition
from a behaviour $\sim s^{-4}$ (diquark form factors equal unity)
to the dimensional counting rule behaviour $\sim s^{-6}$ which is the 
asymptotic result. The other three cross sections, $\sigma_L$,  $\sigma_{TT}$
and  $\sigma_{LT}$, are much smaller than the transverse cross section
in particular around $90^\circ$. Only in the very forward and backward 
directions these cross sections become sizeable but again these regions,
say $|\cos \theta|\ge 0.7$, are outside the hard scattering domain and
the application of a model relying on perturbative QCD is suspect.
In Fig.~\ref{fVC6} the $Q^2$ dependence of the VC cross sections are shown
in a hard scattering situation ($s=10\;{\rm GeV}^2$, $\;\cos \theta=-0.6$)
which is accessible at CEBAF. It can be seen that the transition from real
Compton scattering to large $Q^2$ VC scattering is non-trivial;
a rich structure is predicted. A very interesting phenomenon is the change
of the slope of the transverse cross section.

The full VC scattering contribution to the cross section for the 
electroproduction of photons is obtained by combining the results shown
in Fig.~\ref{fVC5} or \ref{fVC6} according to Eq.~(\ref{d4svc}). 

Next we want to investigate the effect of adding coherently the BH
amplitudes to the VC ones. To that purpose we plot in Fig.~\ref{fBH}
the difference between the full $ep\to ep\gamma$ cross section and the
VC scattering contribution to it divided by the full cross section. The 
question of interest is where are the kinematical regions of small 
BH contaminations allowing to measure the VC process? As can
be seen from Fig.~\ref{fBH} dominance of the VC contributions requires
high energies, small values of $|\cos \theta|$ (the actual values depend
strongly on $Q^2$ and the beam energy $k_{0L}$) and an out-of-plane
experiment, i.~e., a large azimuthal angle ($\phi \ge 50^\circ$). For the
actual CEBAF energy of $6\;{\rm GeV}$ the VC contribution only dominates
in the very backward region and for very small values of $Q^2$.
Outside the regions of VC dominance we expect, according to the diquark
model, strong BH contaminations (see Fig.~\ref{fBH}). We have however not
observed pronounced interference phenomena between the VC and the BH 
contributions like the Coulomb-hadronic interference pattern seen in 
elastic two-body reactions sometimes. 
\section{The electron asymmetry}
\label{eas}
The regions of strong BH contaminations offer an interesting possibility 
to measure the phases of the VC amplitudes relative to those of the 
BH amplitudes. As we explain in the Appendix some of the internal quarks, 
diquarks and gluons may go on mass shell. While these propagator poles are 
integrable they lead to phases of the VC amplitudes. Real as well as 
virtual Compton scattering are the simplest reactions in which, to leading 
order in $\alpha_S$, such phases appear \cite{Far:89a} and are calculable 
perturbatively. Thus, it seems to us, that measuring these phases is a 
very interesting check of the hard scattering approach (with and without 
diquarks).

As a consequence of the phases of the VC amplitudes, the VC contributions 
to the helicity amplitudes of the electroproduction of photons also obtain 
non-trivial phases beyond the phases due to the azimuthal angle dependence 
(see Eq.~(\ref{tvc})). In 
other words, the perturbative phases manifest themselves in the fact that 
the $T$ matrix is not self-adjoint. For the BH process on the other hand, 
$T=T^\dagger$ obviously holds. As we are going to demonstrate information 
on the absorptive part $T\,-\,T^\dagger$ can be obtained from the electron 
asymmetry
\begin{equation}
\label{al}
A_L=\frac{\sigma(+)-\sigma(-)}{\sigma(+)+\sigma(-)}
\end{equation}
where $\sigma(\pm)$ are the (differential) cross section for electroproduction 
of photons with specific helicity of the incoming electron. Since at CEBAF
the electron beam is polarized a measurement of the asymmetry seems
feasible to us. We emphasize that this possibility of measuring the 
absorptive part of the $T$ matrix by the electron asymmetry does not
depend on our model but is quite general and follows from parity and
time reversal invariance merely. As is well-known a one-particle helicity
state transforms under the combined parity and time reversal operation as
\begin{equation}
   |{\bf k},\lambda\rangle \longrightarrow \eta(\lambda)|{\bf k},-\lambda\rangle
\end{equation}
where $\eta(\lambda)$ is $\pm 1$ depending on $\lambda$, the spin of the 
particle and on its internal parity. So the combined
parity and time reversal operation transform a given helicity state into
a state with the same momentum but with reversed helicity. If the interaction
is invariant under the parity and time reversal operation the $T$-matrix
elements for, say, a $2\to3$ body process satisfy the relation
\begin{eqnarray}
\label{tpt}
\langle {\bf k}_1^\prime,\lambda_1^\prime;{\bf k}_2^\prime,\lambda_2^\prime; 
      {\bf k}_3^\prime,\lambda_3^\prime|T|{\bf k}_1,\lambda_1;
      {\bf k}_2,\lambda_2\rangle \hspace{6.0cm}\nonumber\\
\hspace{4.0cm}=\prod_i\eta_i\langle {\bf k}_1^\prime,-\lambda_1^\prime;
       {\bf k}_2^\prime,-\lambda_2^\prime;{\bf k}_3^\prime,-\lambda_3^\prime
       |T^\dagger|{\bf k}_1,-\lambda_1;{\bf k}_2,-\lambda_2;\rangle^\ast.
\end{eqnarray}
(For a two-body amplitude one may also use reflection invariance with
respect to the scattering plane.) Let us now assume that particle 1
is a spin $1/2$ one, say an electron. Then, ignoring all kinematical variables,
the cross section for particle 1 being in a definite helicity state is
\begin{equation}
\sigma(\pm)=\sum_{\{\lambda_i,\lambda_i^\prime\}}|
     T_{\lambda_1^\prime,\lambda_2^\prime,\lambda_3^\prime ;\pm,\lambda_2}|^2.
\end{equation}
The difference of these cross sections may be written as
\begin{eqnarray}
\sigma(+)-\sigma(-)=\Re e\sum_{\{\lambda_i,\lambda_i^\prime\}}
         [T_{\lambda_1^\prime,\lambda_2^\prime,\lambda_3^\prime ;+,\lambda_2}
        +\prod_i\eta_i T_{-\lambda_1^\prime,-\lambda_2^\prime-\lambda_3^\prime;
         -,-\lambda_2}^\ast] \nonumber\\
\hspace{4.0cm} \times  [T_{\lambda_1^\prime,\lambda_2^\prime,\lambda_3^\prime;
               +,\lambda_2}^\ast
     -\prod_i\eta_i T_{-\lambda_1^\prime,-\lambda_2^\prime,-\lambda_3^\prime;
        -,-\lambda_2}].
\end{eqnarray}
If there would be no absorptive part of $T$, i.~e., $T=T^\dagger$, then, 
according to (\ref{tpt}), the difference of the two helicity cross sections
and hence the electron asymmetry (\ref{al}), is zero. Therefore, the asymmtry
measures the non-trivial phase as we claimed above.

Specifying this result to the VC contribution, we find
for the difference of the two cross sections (using the notation of Sect.~II
and dropping the kinematical factors)
\begin{eqnarray}
\sigma(+)-\sigma(-)=-8\sqrt{\frac{\varepsilon}{1-\varepsilon}} \sin \phi
\hspace{8.0cm}\nonumber\\
\hspace{2.0cm}\times\Im m\left[\Phi_9^\ast\,(\Phi_1-\Phi_7)
      +\Phi_{10}^\ast\,(\Phi_2+\Phi_8)
      +\Phi_{11}^\ast\,(\Phi_3-\Phi_5)
      +\Phi_{12}^\ast\,(\Phi_4+\Phi_6)\right]
\end{eqnarray}
Thus, we see that $A_L^{VC}$ measures the imaginary part the 
longitudinal-transverse interference term whereas $\sigma_{LT}$ 
(see Eq.~(\ref{siglt})) measures its real part. In other words, $A_L^{VC}$ 
measures the relative phase between the longitudinal and transverse VC helicity
amplitudes. It turns out, however, that the diquark model predicts
only very small values for $A_L^{VC}$. The ultimate reason is that the 
longitudinal helicity amplitudes are much smaller than the transverse ones
for VC scattering (see Fig.~\ref{fVC5}). So the asymetry is essentially due to the BH-VC interference where  the transverse VC amplitudes do contribute, thereby producing a sizable asymetry.  This asymetry is mostly due to the transverse VC amplitudes because, if we switch off the longitudinal VC amplitudes, the result changes by only a few percent.

Another possibility to measure the non-trivial perturbative phase is
offered by the proton. One may measure the proton asymmetry or the polarization
of the (outgoing) proton in either the electroproduction of photons
or (real or virtual) Compton scattering. For the latter two-body processes
this asymmetry or polarization is related to the imaginary part of the
products of helicity flip and non-flip amplitudes as is well-known.
The predictions from the diquark model for the proton polarization
in the case of real Compton scattering is discussed by Kroll,
Sch\"urmann and Schweiger \cite{Kro:87}.
 
In Figs.~\ref{fA1} we show the electron asymmetry at a beam energy of 
$10\;{\rm GeV}$ and $s=5\;{\rm GeV}^2$ for several values of $Q^2$ and
of the azimuthal angle. $A_L$ is generally very small but in 
regions of strong BH contamination it is sometimes spectacularly enhanced.
Finally in Fig.~\ref{fA2} we present results for the electron asymmetry
for the actual kinematical situation at CEBAF ($s=5\;{\rm GeV}^2$   
$\;k_{0L}=10\;{\rm GeV}^2$,$\;Q^2=1\;{\rm GeV}^2$). As can be seen from
that figure the asymmetry is large in our model for small values of
$|\cos\theta|\;$ ($\simeq 0.2$) and values of the azimuthal angle around
$30^\circ$. The magnitude of the effect is very sensitive to details of 
the model and, therefore, should not be taken literally. Despite of this
one may take our results as an example of what may happen. In so far
we believe that is urgent and important to explore that phenomenon
because it will elucidate strikingly the underlying dynamics of
the electroproduction of photons in a kinematical situation which can be 
considered as a (fairly) hard scattering region.
\section{Concluding remarks}
We have calculated VC scattering off protons within the framework of the
diquark model which represents a particular version of the Brodsky-Lepage model 
appropriate for moderately large momentum transfer. The diquark model
combines perturbative QCD with non-perturbative elements - the diquarks
which represent quark-quark correlations in the proton wave function
modeled as quasi-elementary constituents. Predictions for the VC-scattering
cross section and for the $ep\to ep\gamma$ cross section are presented
for kinematical situations accessible at CEBAF and, perhaps, in the future
at an high energy accelerator like ELFE. It is also shown that the diquark
model predicts a cross section for real Compton scattering in fair
agreement with the data. We have also elaborated on the BH contamination
of the electroproduction of photons which, according to the diquark model,
become sizeable for small azimuthal angles. The BH contribution offers
also the interesting possibility of measuring the relative phases
between the VC and the BH amplitudes. The phases of the VC amplitudes are
a non-trivial phenomenon generated by the fact that some of the internal
quarks, diquarks and gluons may go on mass shell (note that the ingoing
and outgoing quarks and diquarks forming the protons are always on
mass shell). The electron asymmetry $A_L$ is particularly sensitive
to the relative phases and can in principle be measured at CEBAF.
The polarization of the outgoing nucleon in (real as well as in
virtual) Compton scattering is also a measure for the relative phases
between flip and non-flip helicity amplitudes. 

A few words about the self-consistency of our calculations within
perturbative QCD are in order. As can be seen from (\ref{vir}) the running
coupling constant $\alpha_s$ diverges in the end-point regions 
$x_1, y_1 \to 0,1$. This is not characteristic of the diquark model, the 
same happens in the pure quark HSA. As a consequence, perturbation theory
looses its meaning as a weak-coupling expansion. This has previously been 
pointed out by Isgur and Llewellyn-Smith \cite{Isg:89} in their analysis of 
form factors. One way out - actually the one we employ - is to freeze the 
running coupling constant once it has reached a certain value ($0.5$ in the 
case at hand) \cite{Cor:82}. In addition we make use of DAs which strongly 
suppress the end-point regions. The freezing of $\alpha_s$ introduces a new 
external parameter. The modified HSA proposed by Sterman and Li \cite{Li:92a}, 
in which the transverse momenta of the constituents as well as Sudakov 
corrections are taken into account, avoids the introduction of such an external
parameter. The Sudakov corrections select components of the wave functions with
small spatial separations of the constituents. With increasing momentum 
transfer the allowed spatial separations are getting smaller and smaller. 
Asymptotically, only those components remain which are taken into account in 
the standard HSA of Brodsky and Lepage \cite{Bro:80}. The numerical effect of 
the Sudakov suppressions is similar to that of freezing in $\alpha_s$. This has
been demonstrated recently in \cite{Li:92a,Li:92b,Bol:94,Jak:93a} for 
electromagnetic form factors (within the pure HSA). It is claimed that in this 
way the self-consistency of the perturbative contributions can be reestablished
for $Q$ larger than a few GeV without introducing a freezing parameter. We also
note that the transverse momentum dependence of the hadronic wave function 
helps in achieving self-consistency at rather low values of momentum transfer 
\cite{Jak:93a}. In \cite{Jak:93b} an analogous calculation for the S-diquark 
contributions to the proton form factor has been performed and the authors of 
\cite{Jak:93b} arrive at the same conclusion. A treatment of the end-point
regions in the manner proposed by Li and Sterman \cite{Li:92a} diminishes
the results for the form factors obtained with the diquark model by only 
$10$ to $20\%$. Since Sudakov corrections mainly depend on colour and not 
on spin a similar behaviour for $V$ diquarks is to be expected. That small
suppression can be compensated for by adjusting the parameters of the
diquark model appropriately. Effects of similar magnitude are expected
for Compton scattering. We consider these suppressions not explicitely
but understand them as being absorbed in our parameters. For an effective
model as the diquark model is, this procedure is sufficient.\\
\newpage
\appendix 
\section{Helicity amplitudes for 
$\gamma^\ast\,\lowercase{q}\,S\rightarrow \gamma\,\lowercase{q}\,S$}
\label{aha}
The process $\gamma^\ast\,q\,S\rightarrow \gamma\,q\,S$ only contributes
to the hadronic helicity-conserving amplitudes. This implies that 6 of the 12 
independent amplitudes $\Phi_i^{(S,n)}$ given in (\ref{helcs}) are zero
\begin{equation}
\Phi_2^{(S,n)}=\Phi_4^{(S,n)}=\Phi_6^{(S,n)}=\Phi_8^{(S,n)}
=\Phi_{10}^{(S,n)}=\Phi_{12}^{(S,n)}=0\;.
\end{equation}
For dynamical reasons the diquark model yields also
\begin{equation}
\Phi_7^{(S,n)}=0\;.
\end{equation}
The remaining amplitudes $\Phi_i^{(S,n)}$ read in our model:
\begin{eqnarray}
&&
\Phi_1^{(S,3)}=\frac{4\, A_T^{(S,3)}}{q_1^2+i\varepsilon}\frac{s^2}{t\,u(s+Q^2)}
\left\{\frac{x_1 s}{x_2 y_2}-\frac{1}{y_1 u-y_2 Q^2}
\left(\frac{u\, Q^2}{y_1 y_2}-(u+Q^2)^2\right)\right\}\nonumber\\
&&
\Phi_1^{(S,4)}=4\, A_T^{(S,4)}\frac{1}{u}
\left\{\frac{(s+Q^2)}{q_1^2+i\varepsilon}\frac{(s+x_2t)}{g_{4a}^2+i\varepsilon}
-\frac{1}{g_{4b}^2+i\varepsilon}\left(t+\frac{u}{y_1}\right)\right\}\nonumber\\
&&
\Phi_1^{(S,5)}=4\, A_T^{(S,5)}\frac{1}{x_1 y_1 u}\\ \nonumber \\ 
&&
\Phi_3^{(S,3)}=\frac{4 A_T^{(S,3)}}{y_1 u-y_2 Q^2}\frac{Q^2}{u}
\left\{\frac{y_1 u}{x_2 y_2(s+Q^2)}+\frac{1}{q_1^2+i\varepsilon}
\left(s+Q^2-\frac{s\,Q^2}{x_1 x_2(s+Q^2)}\right)\right\} \nonumber\\
&&
\Phi_3^{(S,4)}=\frac{4\, A_T^{(S,4)}}{y_2 u-y_1 Q^2}\frac{t\,Q^2}{u (s+Q^2)}
\left\{\frac{1}{x_1}+\frac{u+y_2t}{g_{4a}^2+i\varepsilon}\right\}\nonumber\\
&&
\Phi_3^{(S,5)}=-\frac{4\, A_T^{(S,5)}}{y_2 u-y_1 Q^2}
\frac{Q^2(s+Q^2+y_2 t)}{x_1 y_1 u (s+Q^2)}\\ \nonumber \\ 
&&
\Phi_5^{(S,3)}=-\frac{4\, A_T^{(S,3)}}{y_1 u-y_2 Q^2}\frac{s}{t}
\left\{\frac{y_1 u}{x_2 y_2(s+Q^2)}+\frac{1}{q_1^2+i\varepsilon}
\left(s+Q^2-\frac{s\,Q^2}{x_1 x_2(s+Q^2)}\right)\right\}\nonumber\\
&&
\Phi_5^{(S,4)}=\frac{4\, A_T^{(S,4)}}{y_2 u-y_1 Q^2}
\left\{-\frac{s}{x_1(s+Q^2)}-\frac{(u+y_2t)}{s+Q^2}
\frac{s}{g_{4a}^2+i\varepsilon}
+\frac{x_2 s}{x_1}\frac{1}{g_{4b}^2+i\varepsilon}\right\}\nonumber\\
&&
\Phi_5^{(S,5)}=\frac{4\, A_T^{(S,5)}}{y_2 u-y_1 Q^2}
\frac{y_2 s}{x_1 y_1 u (s+Q^2)}\\ \nonumber \\
&&
\Phi_9^{(S,3)}=0\nonumber \\
&&
\Phi_9^{(S,4)}=2\, A_0^{(S,4)}\frac{y_2 s}
{(g_{4a}^2+i\varepsilon)(D_1^2+i\varepsilon)}\nonumber \\
&&
\Phi_9^{(S,5)}=-2\,A_0^{(S,5)}\frac{s}{x_1 y_1 t}\frac{1}
{D_1^2+i\varepsilon}\\ \nonumber \\
&&
\Phi_{12}^{(S,3)}=-\frac{4\, A_0^{(S,3)}}{y_1 u-y_2 Q^2}\frac{s}{t}
\left\{\frac{y_1 u}{x_2 y_2(s+Q^2)}+\frac{1}{q_1^2+i\varepsilon}
\left(s+Q^2-\frac{s Q^2}{x_1 x_2(s+Q^2)}\right)\right\}\nonumber\\
&&
\Phi_{12}^{(S,4)}=\frac{2\, A_0^{(S,4)}}{y_2 u-y_1 Q^2}
\left\{-\frac{2 s}{x_1(s+Q^2)}+2\left(1+\frac{y_1 t}{s+Q^2}\right)
\frac{s^2}{g_{4a}^2+i\varepsilon}+\frac{x_2 s}{x_1}
\frac{1}{g_{4b}^2+i\varepsilon}\right\}\nonumber\\
&&
\Phi_{12}^{(S,5)}=\frac{2\, A_0^{(S,5)}}{y_2 u-y_1 Q^2}\frac{s}{t}
\frac{s+Q^2+2\,y_2 t}{x_1 y_1(s+Q^2)}
\end{eqnarray}
where 
\begin{eqnarray}
&&A_T^{(S,n)}=\sqrt{-\frac{u}{s}}\,\alpha_s(g_n^2)\,F_S^{(n)}(Q_n^2)\nonumber\\
&&A_0^{(S,n)}=\frac{\sqrt{-2\,t\,Q^2}}{s}\,\alpha_s(g_n^2)\,F_S^{(n)}(Q_n^2).
\end{eqnarray}
For $Q^2=0$ the $\Phi_3^{(S,n)}$ are zero and the remaining 
amplitudes agree with those quoted by Kroll, Sch\"urmann and 
Schweiger \cite{Kro:87}. The various gluon propagators appearing as 
arguments in $\alpha_s$ and as arguments in the various diquark vertex 
functions are taken as:
\begin{equation}
\label{vir}
\begin{array}{lll}
g_3^2=-x_2 y_2 t\quad & g_4^2=-\frac{1}{2}(x_1 y_2+x_2 y_1)t\quad 
                                                   &g_5^2=-x_1 y_1 t \\ 
[0.2cm]
Q_3^2=g_3^2 &Q_4^2=g_4^2+Q_3^2 & Q_5^2=g_5^2+Q_3^2
\end{array}
\end{equation}
$g_{4a},\,g_{4b},\,q_1$ and $D_1$ are the momenta of the internal gluon, quark 
and diquarks which lead to the poles within the range of integration. The 
corresponding virtualities are
\begin{equation}
\label{pos}
\begin{array}{ll}
q_1^2=x_1 s-x_2 Q^2 &\quad g_{4a}^2=x_1 y_2 s+x_2 y_1 u-x_2 y_2 Q^2\\
D_1^2=x_2 s-x_1 Q^2 &\quad g_{4b}^2=x_2 y_1 s+x_1 y_2 u-x_1 y_1 Q^2.
\end{array}
\end{equation}
The propagator poles are integrable and do not destroy the validity of the 
hard scattering  approach \cite{Far:89}. The poles are handled in the usal 
way by using the $i\varepsilon$ prescription
\begin{equation}
\label{pval}
\frac{1}{x\pm i\epsilon}=P\left(\frac{1}{x}\right)\mp i\pi\delta(x)
\end{equation}
In the case that there is only one pole in the region of integration, the 
integrals over $x_1$ can be performed in the following way:
\begin{eqnarray}\label{Eprop}
&&\int_{0}^{1}dx_1\frac{f(x_1,y_1,s,t,Q^2)}{p^2+i\varepsilon}\phi_D(x_1)
\nonumber \\
&&
=\int_{0}^{1}dx_1
\frac{f(x_1,y_1,s,t,Q^2)\,\phi_D(x_1)
     -f(x_1^{(p)},y_1,s,t,Q^2)\,\phi_D(x_1^{(p)})}{p^2} \\
&&+
f(x_1^{(p)},y_1,s,t,Q^2)\,\phi_D(x_1^{(p)})\,
\left(P.V.\int_{0}^{1}\frac{dx_1}{p^2}
-i\,\pi\,{\left|\frac{\partial p^2}{\partial x_1}\right|}^{-1}\right).\nonumber
\end{eqnarray}
where $x_1^{(p)}$ is the zero of the equation $p^2=0$ and the different 
$p^2$ are given in Eq.~(\ref{pos}). The necessary principal value integral 
over $x_1$ is then carried out analytically. The different integrals 
and the derivatives of the propagators are listed in Table I. 
In cases when two propagators go on shell at the same time 
one has to make a partial fractioning before applying (\ref{Eprop}). 

\newcommand{\D}{\displaystyle}
\begin{table}
\label{tp}
\caption[]{\label{tsing}Zeros of propagators from Eq.~(\ref{Eprop}), 
partial derivatives and principal value integrals.}
\begin{tabular}{cccc}
$ p^2 $&$ x_1^{(p)} $&$\D\left|\frac{\partial p^2}{\partial x_1}\right| $&
$\D P.V.\int_{0}^{1}\frac{dx_1}{p^2} $ \\[0.5ex] \tableline
  & & &  \\
$ q_1^2 $&$\D \frac{Q^2}{s+Q^2} $&
$ s+Q^2 $&$\D \frac{1}{s+Q^2}\ln\left(\frac{s}{Q^2}\right) $
\\
  & & &  \\
$ D_1^2 $&$\D \frac{s}{s+Q^2} $&
$ s+Q^2 $&$\D \frac{1}{s+Q^2}\ln\left(\frac{s}{Q^2}\right) $
\\
  & & &  \\
$ g_{4a}^2 $&
$\D \frac{y_1u-y_2Q^2}{y_2t+u} $&
$ -(y_2t+u) $&
$\D \frac{-1}{y_1t+u}
  \ln\left(\frac{y_1u+y_2 Q^2}{y_2t-u}\right) $\\[0.3cm]
$ g_{4b}^2 $&
$\D -\frac{y_1s}{y_1t+u} $&
$ -(y_1t+u) $&
$\D \frac{-1}{y_1t+u}
  \ln\left(\frac{y_1s}{y_1t-u}\right) $\\[0.3cm]
\end{tabular}
\end{table}
\begin{figure}
\caption[]{The Bethe-Heitler and the virtual Compton contributions to the 
process $e\,p\rightarrow e\,p\,\gamma$. The momenta and the helicities of the 
various particles appearing in that process are indicated.}
\label{fkin}
\end{figure}

\begin{figure}
\caption[]{Feynman graphs contributing to the process 
$\gamma^\ast\,p\rightarrow\gamma\,p$. Graphs with the two 
photons interchanged are not shown.}
\label{frfd}
\end{figure}


\begin{figure}
\caption[]{The cross section for real Compton scattering off protons scaled 
by $s^6$ vs.~$\cos\theta$ for three different photon energies. 
The experimental data are taken from \cite{Shu:79}.} 
\label{frcc}
\end{figure}

\begin{figure}
\caption[]{The cross section for virtual Compton scattering vs.~$\cos\theta$ 
for several values of $Q^2/s$ at a) $s=5\;{\rm GeV}^2$, b) $10\;{\rm GeV}^2$. 
Upper left: the transverse cross section scaled by $s^6$. Upper right: the 
ratio of the longitudinal over the transverse cross sections. Lower left 
(right): the ratio of the longitudinal (transverse) - transverse interference 
term over the transverse cross section.}
\label{fVC5}
\end{figure}

\begin{figure}
\caption[]{The cross section for virtual Compton scattering vs.~$Q^2$. 
For notations see Fig.~\ref{fVC5}.}
\label{fVC6}
\end{figure}

\begin{figure}
\caption[]{The difference of the full photon electroproduction cross 
section and the VC contribution to it over the full cross section vs.~$\cos 
\theta$ for several combinations of values of the beam energy $k_{0L}$, 
$Q^2$ and the azimuthal angle $\Phi$. a) $s=5\;{\rm GeV}^2$, 
b) $10\;{\rm GeV}^2$}
\label{fBH}
\end{figure}

\begin{figure}
\caption[]{The electron asymmetry $A_L$ vs.~$\cos \theta$ for several values 
of $Q^2$ and of the azimuthal angle.}
\label{fA1}
\end{figure}

\begin{figure}
\caption[]{The electron asymmetry at CEBAF. Top: $A_L$ vs.~$\phi$ 
for several values of $\cos \theta$. Bottom: $A_L$ vs.~$\cos \theta$ for 
several values of $\phi$.}
\label{fA2}
\end{figure}

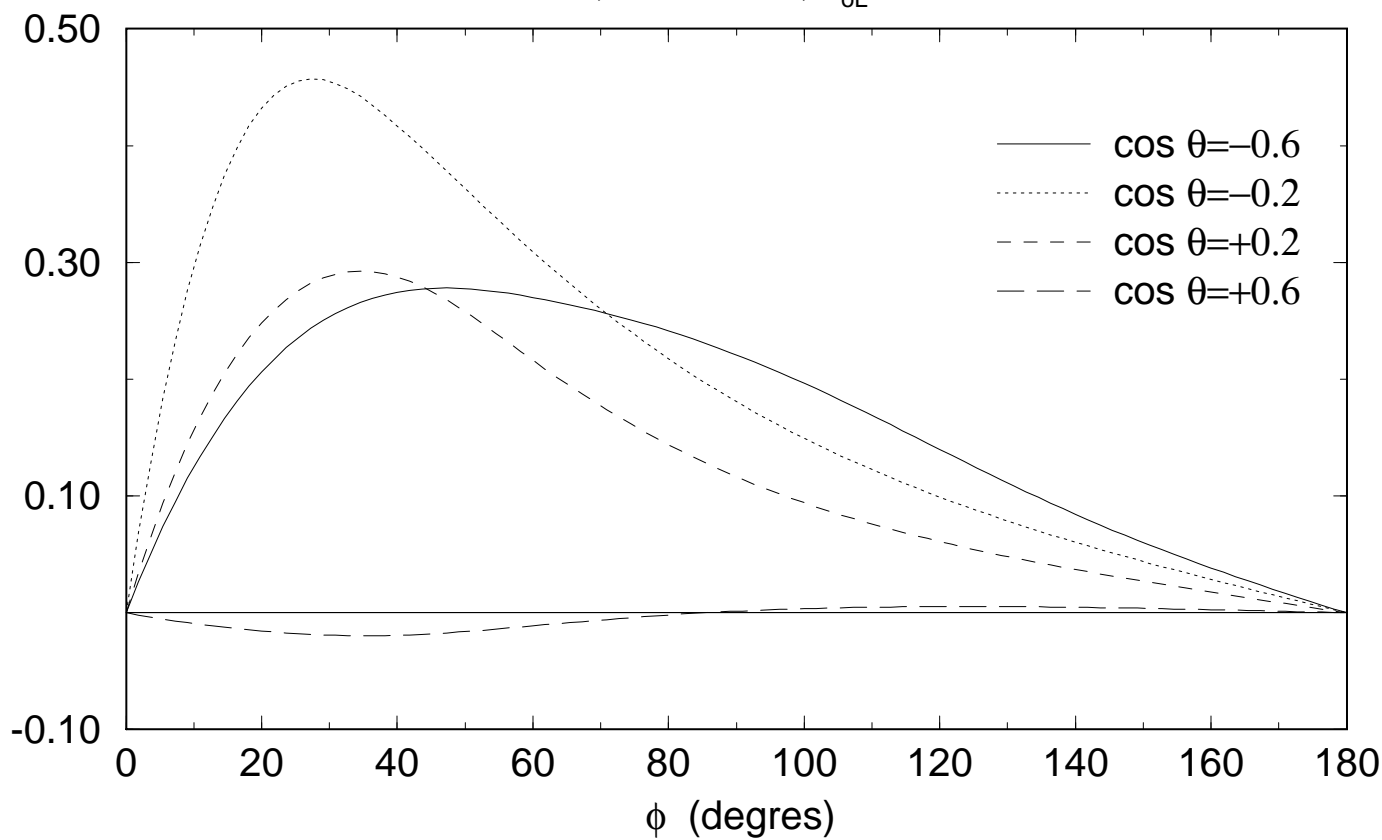
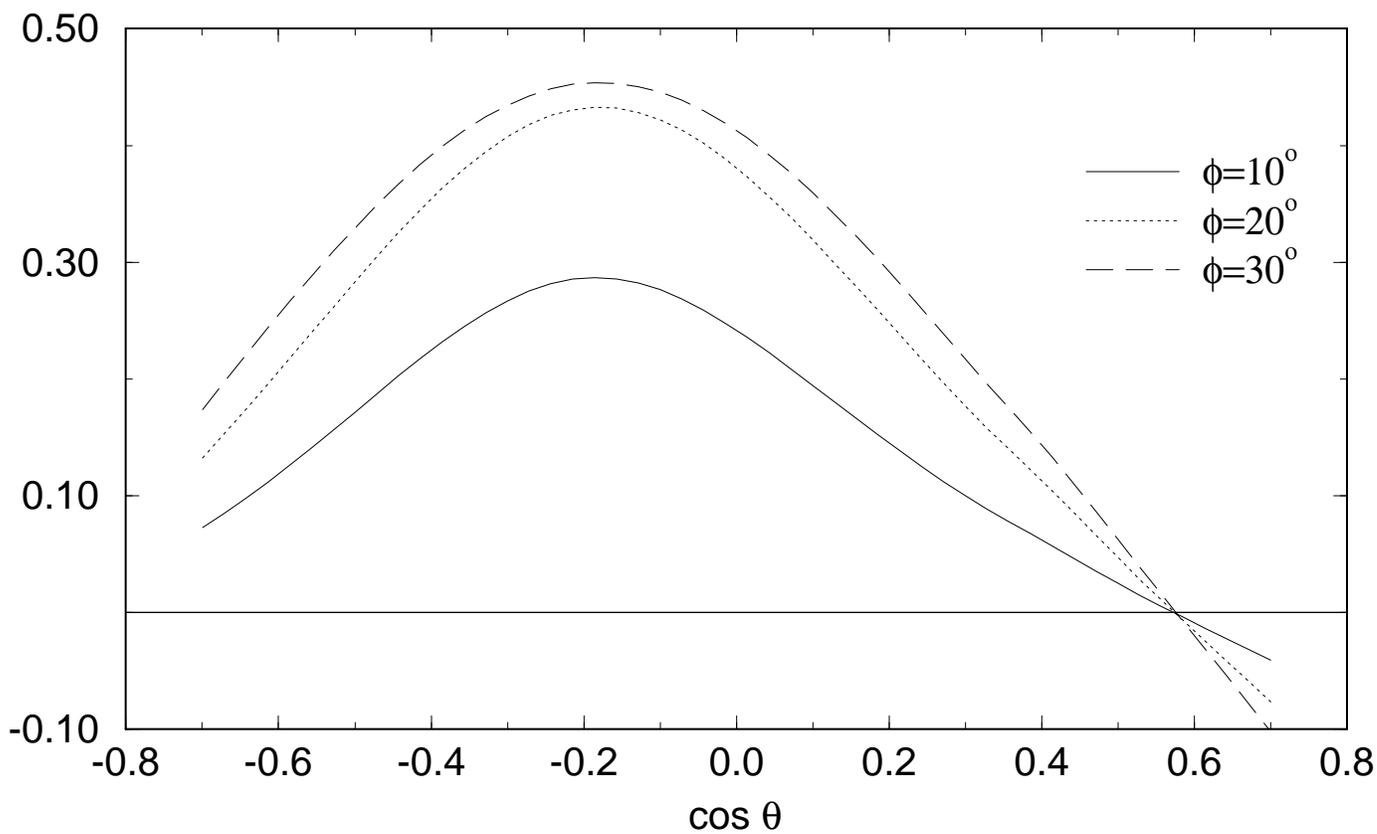
\end{document}